# Cyclic models of the relativistic universe: the early history

Helge Kragh[*]

**Abstract**. Within the framework of relativistic cosmology oscillating or cyclic models of the universe were introduced by A. Friedmann in his seminal paper of 1922. With the recognition of evolutionary cosmology in the 1930s this class of closed models attracted considerable interest and was investigated by several physicists and astronomers. Whereas the Friedmann-Einstein model exhibited only a single maximum value, R. Tolman argued for an endless series of cycles. After World War II, cyclic or pulsating models were suggested by W. Bonnor and H. Zanstra, among others, but they remained peripheral to mainstream cosmology. The paper reviews the development from 1922 to the 1960s, paying particular attention to the works of Friedmann, Einstein, Tolman and Zanstra. It also points out the role played by bouncing models in the emergence of modern big-bang cosmology.

Although the general idea of a cyclic or oscillating universe goes back to times immemorial, it was only with the advent of relativistic cosmology that it could be formulated in a mathematically precise way and confronted with observations. Ever since Alexander Friedmann introduced the possibility of a closed cyclic universe in 1922, it has continued to attract interest among a minority of astronomers and physicists. At the same time it has been controversial and widely seen as speculative, in part because of its historical association with an antireligious world view. According to Steven Weinberg, "the oscillating model … nicely avoids the problem of Genesis" and may be considered philosophically appealing for that reason (Weinberg 1977: 154). In spite of many problems and a generally bad reputation, cyclic models never vanished from the scene of cosmology. Indeed, they have recently experienced a remarkable revival, especially

---

[*] Centre for Science Studies, Department of Physics and Astronomy, Aarhus University, Aarhus, Denmark. E-mail: helge.kragh@ivs.au.dk. This is a revised version of a paper prepared several years ago for volume 13 of the *Einstein Studies* series. It is unknown if the volume will ever be published.

in forms inspired by string theory (Steinhardt & Turok 2007; Lehners, Steinhardt & Turok 2009).

The present essay covers the history of the cyclic universe, understood as a class of solutions to the cosmological field equations, in the period from 1922 to the 1960s. No attempt is made to extend the investigation to the later development (which is covered in part in Kragh 2009). As the history of this kind of cosmological view goes up to the present, so it goes very far back in time, if more as a philosophical than a scientific idea. This earlier history is not part of my essay either, but contrary to the modern history it is thoroughly described in the literature of science and ideas (Jaki 1974; Kragh 2008).

## 1. The Friedmann-Einstein Universe

In his seminal work of 1922 which marks the very beginning of evolutionary relativistic cosmology, Friedmann analyzed from a mathematical perspective the various solutions to Einstein's field equations, including the possibility of a zero or negative cosmological constant (Friedmann 1922). His analysis rested on very general assumptions, namely that the cosmic matter is at rest and exerts no pressure, and also that the constant-curvature space is orthogonal to time. What would later be known as the Friedmann equations, appeared in the form

$$\left(\frac{R'}{R}\right)^2 + \frac{2RR''}{R^2} + \frac{c^2}{R^2} - \Lambda = 0$$

and

$$3\left(\frac{R'}{R}\right)^2 + \frac{3c^2}{R^2} - \Lambda = \kappa c^2 \rho \, ,$$

where $\kappa = 8\pi G/c^2$ and $\rho$ denotes the mean density of matter. Introducing a constant $A$ given by the total mass $M$ of the universe according to

$$A = \frac{\kappa M}{6\pi^2} = \frac{4GM}{3\pi c^2} \, ,$$

he showed that, in the case $\Lambda < 4c^2/9A^2$, the radius of curvature would become a periodic function of $t$ with a "world period" given by

$$t_\pi = \frac{2}{c} \int\limits_0^{x_0} \sqrt{\frac{x}{A - x + \frac{\Lambda x^3}{3c^2}}} \, dx$$

Friedmann commented: "The radius of curvature varies between 0 and $x_0$. We shall call this universe the *periodic world*. The period of the periodic world increases if we increase $\Lambda$ and tends towards infinity if $\Lambda$ tends towards the value $\Lambda_1 = 4c^2/9A^2$" (Friedmann 1922: 385).[1] He further noted that for small values of $\Lambda$, the world period is given by

$$t_\pi \cong \frac{\pi A}{c} = \frac{\kappa M}{6\pi c}$$

As an illustration, he calculated that for $\Lambda = 0$ and a world mass $M = 5 \times 10^{21}$ solar masses, $t_\pi$ became of the order of 10 billion years.[2] In his paper of 1922, Friedmann did not describe the universe as oscillating in the sense that cycle followed after cycle, but only referred to a single cycle from bang to crunch. Nor did he refer to thermodynamic or other physical properties such as the content of matter and radiation. Moreover, there was no mention of the galactic redshifts known at the time. His work was basically a mathematical investigation.

It is worth noting that Friedmann, after having introduced the idea of a world cycle, pointed out that the notion could be understood in two different ways. Two events could be counted as coincident if they have the same spatial coordinates at times $t'$ and $t' \pm nt_\pi$ ($n = 1, 2, \ldots$), which corresponds to the ordinary picture of a pulsating universe limited in time between $t = 0$ and $t = t_\pi$. Alternatively, "if the time varies between $-\infty$ and $+\infty$ (e.g. if we consider two events as coincident only when not only their spatial but also their world coordinates coincide), we come to a real periodicity of the space curvature" (Friedmann 1922: 385). Without elaborating, he adopted the first viewpoint. The second one involves the strange notion of a "cyclic time" in a strict sense, where $R$ does not vary periodically in time but time itself moves, as it were, on a circle. When the universe changes from expansion to contraction the direction of time (as given by the entropy, for example) will change as well, implying that the final singularity coincides with the initial one. This kind of cyclic time has been discussed by philosophers, but it has not found any use in science and is generally thought to be absurd (Whitrow 1980: 39-41).

---

[1] For the cosmological constant I have substituted the symbol $\Lambda$ for Friedmann's $\lambda$ (which is also the symbol Einstein used in his article of 1917 in which he introduced the constant). An English translation of Friedmann's paper appears in several versions, e.g. Lang & Gingerich 1979: 838-843.

[2] It is unknown why he chose this particular value of $M$, but see Tropp, Frenkel & Chernin 1993: 159, and also the comment by Georg Singer in Friedmann 2000: 137-138.

Although Friedmann did not express either physical or philosophical preferences for a particular world model, he seems to have been fascinated by the possibility of a periodic or cyclic universe. In a semipopular book published in Russian in 1923, *The World as Space and Time*, he elaborated on the subject:

> Cases are also possible when the radius of curvature changes periodically. The universe contracts into a point (into nothing) and then increases its radius from the point up to a certain value, then again diminishes its radius of curvature, transforms itself into a point, etc. This brings to mind what Hindu mythology has to say about cycles of existence, and it also becomes possible to speak about "the creation of the world from nothing," but all this should at present be considered as curious facts which cannot be reliably supported by the inadequate astronomical material.[3]

As is well known, Friedmann's seminal works of 1922 and 1924 were ignored by contemporary physicists and astronomers, who also failed to pay attention to his novel conception of a cyclic and closed universe. Indeed, his work is a prime example of what is known as a premature discovery (Hetherington 2002).

Three years after Friedmann's paper in the *Zeitschrift für Physik* there appeared in the same journal a lengthy article by the Hungarian physicist Cornelius Lanczos, one of the pioneers in the early phase of relativistic cosmology and also a contributor of some significance to the early development of quantum mechanics. This article too was ignored by contemporary physicists – during the 1920s it received only a single citation – and neither has it been noticed by historians of physics and cosmology. In spite of its lack of impact, it deserves attention.

Without citing Friedmann, Lanczos investigated a world model which was not only closed in space but also in time, with time being constructed as a periodic coordinate (Lanczos 1925). Although he did not refer to Friedmann, I consider it probable that he read the paper of 1922 and received some inspiration from it. As a theoretical cosmologist and a frequent contributor to the *Zeitschrift*, it is hard to believe that he failed to pay attention to Friedmann's work. For one thing, the paper of 1925 adopts a language close to that used by Friedmann. Not only did

[3] Friedmann 2000: 109. The book was translated into French in 1997 (Luminet 1997: 99-214) and into German three years later (Friedmann 2000), introduced and annotated by Georg Singer. An English translation is still missing.

Lanczos speak of a "world period" (*Weltperiode*), the very term introduced by Friedmann, he also distinguished between space varying periodically in time and time being a cyclic parameter. He argued that the latter concept, which leads to an eternally recurring universe in the sense of Friedrich Nietzsche (whom he quoted), was contradictory and of no scientific use. Whatever his inspiration from Friedmann, Lanczos found for the world period $T$ an expression in terms of Planck's constant ($h$), the mass of the electron ($m$), and the radius of the static Einstein universe ($R$):

$$T = \frac{4\pi^2 m}{h}\, R^2 \cong 10^{41} \text{ years}$$

He noted that the number was enormously greater than $R/c$, by a factor of about $10^{32}$. However, the period $T$ was not the period of an oscillating universe, for according to Lanczos his model was not really periodic in time. It is unclear to me how he understood his picture of what may appear to be a closed world varying in time with an enormous world period. The aim of his paper was not to suggest a new cosmological theory, but to relate the quantum phenomena of the microcosmos to the structure of the macrocosmos, namely, to understand Planck's quantum of action in terms of cosmology. "The solution to the quantum riddles is hidden in the spatial and temporal closeness of the universe," he wrote (Lanczos 1925: 80).

   After having learned about Edwin Hubble's analysis of the measurements of nebular redshifts, Einstein abandoned his previous insistence on the static cosmological solution and accepted the expanding universe as a superior alternative. In the spring of 1931, shortly before Lemaître published his idea of an exploding universe, Einstein belatedly recognized Friedmann's pioneering work – which, he wrote, was "uninfluenced by observations" – and decided that there no longer was any need for the cosmological constant. He discussed a Friedmann cyclic model filled with pressureless matter ("dust") and containing no radiation, but without extending it to possible previous or later cycles.[4] That is, his model of 1931 was not oscillatory in the strict sense. From Friedmann's first equation, Einstein obtained the expression

---

[4] Steinhardt and Turok 2007: 177 suggest that Einstein's choice of investigating the periodic Friedmann solution was a result of his "philosophical predilections" and fascination of Spinoza's philosophy. The suggestion lacks documentary evidence as well as plausibility. For a positive evaluation of Einstein's "cycloidal universe," see Bonnor 1964: 96-101.

$$\left(\frac{dP}{dt}\right)^2 = c^2 \frac{P_0 - P}{P},$$

where $P = P(t)$ is the curvature radius of the closed universe and $P_0$ its maximum value. He explained: "For small $P$ (our idealization is invalid for the strict limit $P = 0$) $P$ increases very rapidly. Then, as $P$ increases, the speed of change $dP/dt$ decreases ever more and becomes zero at the limiting value $P = P_0$, after which the entire process takes place in the opposite sense (that is, with $P$ decreasing at an increasing speed)" (Einstein 1931: 237). Assuming that $P - P_0$ was of the same magnitude as $P_0$ it followed from his rewriting of the second of Friedmann's equations that

$$D^2 \cong \frac{1}{3}\rho\kappa, \quad \text{with} \quad D = \frac{1}{P}\frac{dP}{dt}\frac{1}{c}$$

The quantity $cD$ is the Hubble constant, a name Einstein did not use. Disregarding the factor 1/3 he obtained as a "mere order of magnitude" a mean density of matter as high as $10^{-26}$ g/cm³ and from $P \sim 1/D$ a current world radius of only about 100 million light years. He did not state the value of $H = cD$.

At the end of the paper Einstein summarized what he considered to be its significance:

> This theory is sufficiently simple that it can be conveniently compared with the astronomical data. It further shows how cautious one should be with large extrapolations of the time in astronomy. It is, first of all, remarkable that the general theory of relativity seems able to justify in a more natural way (namely, without the $\Lambda$ term) Hubbel's [sic] new data than the postulate of the quasi-static nature of space, which now has little empirical support.[5]

The 1931 paper is not among Einstein's better known works. Yet it is noteworthy, and that not only because it marked Einstein's public retraction of the cosmological constant, but also because he explicitly formulated a version of what soon became known as the cosmological principle: "All places in the universe are equivalent [*gleichgültig*]." Moreover, formally the model belonged to the big-bang class, indeed it was the first model ever of this kind.

Later the same year Richard Tolman investigated Einstein's model in greater detail, in particular by introducing thermodynamical considerations. By using relativistic thermodynamics he obtained the surprising result that the expansion and contraction of the model universe were not accompanied by increase in entropy, from which he concluded that they "could presumably be repeated over and over again" (Tolman 1931b: 1761). Tolman also provided a general expression of the way in which the radius of the Einstein cyclic universe varied in time. His result was

$$\sqrt{\frac{R}{R_m}} = \sin\left(\frac{t}{R_m} + \sqrt{\frac{R}{R_m}\left(1 - \frac{R}{R_m}\right)}\right),$$

where $R_m$ is a constant signifying the upper limit of $R$. The radius would expand from $R = 0$ at $t = 0$ to $R_m$ at $t = \pi R_m/2$ and then return to zero at $t = \pi R_m$. Written in a parametric form (Tolman 1934: 413) the expression represents a cycloid in the $Rt$ plane given by

$$R = \frac{\alpha}{3}(1 - \cos\psi), \quad t = \frac{\alpha}{6}(1 - \sin\psi),$$

where $\alpha$ denotes the constant quantity $8\pi\rho R^3$. The radius will oscillate between $R = 0$ and $R_m = \alpha/3$ at $t = \pi\alpha/6$.

---

[5] Einstein systematically misspelled Hubble's name as "Hubbel." As noted in Nussbaumer & Bieri 2009, this may indicate that he had not actually read Hubble's papers but only been told about them.

The cyclic or pulsating model Einstein proposed in 1931 held no special significance for him, such as shown by the model he developed the following year in collaboration with Willem de Sitter (Einstein & de Sitter 1931). In the well-known Einstein-de Sitter model the pressure and the cosmological constant were assumed to be zero, as in the earlier model, but it also assumed a flat space and consequently was steadily expanding according to

$$R(t) = at^{2/3} + b$$

In Einstein's view, the significance of his papers of 1931 and 1932 was not so much that they described new cosmological models, but that they demonstrated that the cosmological constant was unnecessary. This was an "incomparable relief," as he wrote to Tolman in the summer of 1931, including with his letter a copy of his paper in the proceedings of the Prussian Academy. Einstein further pointed out the difficulty with the singularities formally appearing at $t = 0$ and $t = t_{max}$, suggesting that they might disappear in a more realistic version of the model. Tolman responded:

When I first saw your proposed quasi-periodic solution for the cosmological line element, I was very much troubled by the difficulties connected with the behavior of the model in the neighborhood of the points of zero proper volume. The remarks in your letter, however, pointing out that the actual inhomogeneity in the distribution of matter might make the idealized treatment fail in that neighborhood, seem to me very important. … I think that it is pertinent to remark that from a physical point of view contraction to a very small volume could only be followed by renewed expansion. Hence all in all I am feeling much more comfortable about this difficulty, and indeed have just sent an article to the Physical Review discussing among other things the application of relativistic thermodynamics to quasi-periodic models of the universe.[6]

At the end of his paper of 1931 Einstein noted the much discussed time-scale difficulty, namely that the time allowed by the cosmological model – he stated it to be about 10 billion years – was much smaller than the age of the stars and galaxies as estimated at the time (Kragh 1996: 73-79). This was a serious problem in the Einstein-de Sitter model, where $t = 2T_0/3$ with $T_0$ the Hubble time, which in the 1930s was believed to be about 1.8 billion years. The problem was even more serious in the denser oscillating models, where the present age must be less than 1.2 billion years. Einstein suggested that "one can try to get out of the difficulty by pointing out that the inhomogeneity of stellar matter makes our approximate treatment illusory" (Einstein 1931: 237).

In a later survey of the cosmological problem, first published in 1945, he repeated his suggestion that the theory was "inadequate for very high density of matter" (Einstein 1953: 124; Einstein 1945). Although Howard Percy Robertson at Princeton University did not endorse Einstein's model, he agreed that it was "emotionally more satisfactory" to assume that the field equations break down near $R = 0$ and leave room for a non-singular bounce (Robertson 1932: 224).

Unbeknownst to Einstein, a physicist from Japan had investigated cyclic models of the universe a little earlier than himself. In September 1930 the Japanese theoretical physicist Tokio Takeuchi read a paper to the Physico-Mathematical Society of Japan on the cyclic universe which was published the following year (Takeuchi 1931).[7] Apparently unaware of Friedmann's earlier work, Takeuchi constructed a complicated cyclic line element which he claimed was "in agreement with the view of Boltzmann."[8] From a philosophical point of view he found a monotonically increasing universe to be "not pleasing." His theory had the advantage not only of securing the eternity of the universe but also of avoiding singularities where the energy-momentum tensor becomes infinite. For the volume of the oscillating universe he found the expression

$$V(t) = \exp\left(\frac{3}{2}\sin kt\right) 2\pi^2 R^3,$$

where $k$ and $R$ are constants and the speed of light is taken as unity. The universe will thus reach a maximum size at $kt = \pi/2$. Inspired by Tolman, he discussed the thermodynamical properties of his cyclic model universe, including its brightness and the transformation of matter into energy. Published in a not widely circulated Japanese journal, Takeuchi's theory attracted almost no attention. It was however noticed by Tolman, who dismissed it as artificial and devoid of physical interest (Tolman 1931b: 1764).

---

[6]  Tolman to Einstein, 14 September 1931, a response to Einstein to Tolman, 27 June 1931. Courtesy the Einstein Archives and Princeton University Press. Tolman's reference to his forthcoming paper was to Tolman 1931b. See also Nussbaumer & Bieri 2009: 145-149, where relevant fragments of Einstein's diary from the spring of 1931 are cited.

## 2. A controversial universe

The time-scale problem was not only a concern of Einstein's, it also worried de Sitter who for a time thought it justified a kind of pulsating universe, although not in Einstein's sense. The Dutch astronomer speculated that the universe may have "shrinked during an infinite time from an infinite radius to a minimum value, … increasing again afterwards, the minimum being reached a few thousand million years ago" (de Sitter 1931a: 7).[9] However, he realized that there was no very good reason to advocate such a cosmic scenario. In fact, in a systematic study of cosmological solutions from the summer of 1931 he concluded that all oscillating models were ruled out as incompatible with empirical data (de Sitter 1931b).

---

[7] Takeuchi wrote several papers on relativity, quantum theory and cosmology in the years around 1930, some of them in the proceedings of the Physico-Mathematical Society and others in the *Zeitschrift für Physik*. For example, he investigated the hypothesis of a decreasing velocity of light within the framework of evolutionary cosmology, concluding that the decrease was only about 1 cm/s/year (Takeuchi 1930).

[8] In fact, Boltzmann never advocated or discussed an oscillating universe. In 1895 he developed a remarkable scenario of a kind of multiverse, including "worlds" with a reversed entropic order, but he did not consider a series of such worlds changing periodically in time (Boltzmann 1895).

[9] A similar speculation appeared in Robertson (1932: 224), who suggested that the universe "was originally shrinking and, having reached a finite lower limit, began to expand."

A somewhat similar critique came from Robertson in his influential review of cosmological models in the *Reviews of Modern Physics* from 1933. Robertson pointed out that Einstein's pulsating model required an unrealistically high matter density, namely about $10^{-27}$ g/cm$^3$, and that this was several thousand times more than indicated by observations (Robertson 1933: 78). His own favorite model was the Lemaître-Eddington solution, where the universe expands gently and monotonically from an unstable Einstein state.

In a careful study of Friedmann-Lemaître world models, Gawrilow Raschco Zaycoff, a Bulgarian physicist and physics teacher, commented on the repeated "births" and "deaths" of the oscillating universe. An examination of the possibility of a lower limit $R > 0$ led him to conclude that, irrespective the value of the cosmological constant, "there exist no periodic solutions to the gravitational equations of the cosmological problem" (Zaycoff 1933: 135). However, he left open the possibility of non-singular bounces in the case of a modification of the field equations.

De Sitter continued to speculate that considerations of a possible state of the universe before $R = 0$ might solve the time-scale difficulty. At a meeting of the Royal Astronomical Society in 1933 he suggested that perhaps the universe had once contracted to a point-like state, with all the galaxies passing simultaneously through it some 3-5 billion years ago. By assuming that the stars had survived this "very vigorous" critical event, their true ages could be much longer than the age of the universe as based on the recession of the galaxies (de Sitter 1933a: 184). De Sitter believed that observations favored either a steadily expanding universe of the Einstein-de Sitter type or a bouncing universe. He admitted that the alternative of an oscillating universe did not appeal to him: "Personally I have, like Eddington, a strong dislike to a periodic universe, but that is a purely personal idiosyncracy, not based on any physical or astronomical data" (de Sitter 1933b: 630).

Eddington not only disliked Lemaître's hypothesis of an ever-expanding universe with an origin a finite time ago, he was equally opposed to the idea of a cyclic universe, whether in its classical or relativistic version. In agreement with the critique raised against the cyclic universe in the nineteenth century, he found the idea to contradict the law of entropy increase, which he considered absolutely fundamental. "I am no Phoenix worshipper," he admitted (Eddington 1928: 86):

> I would feel more content that the Universe should accomplish some great scheme of evolution and, having achieved whatever may be achieved, lapse back into chaotic changelessness, than its purpose should be banalized by continual repetition. I am an Evolutionist, not a Multiplicationist. It seems rather stupid to keep doing the same thing over and over again.

The change from the static to the evolving universe did not cause Eddington to change his view. Neither did Tolman's revision of the thermodynamics of the universe, which he ignored. His reason for dismissing the cyclic universe was not primarily scientific, but rather based in religious and moral sentiments: "From a *moral standpoint* the conception of a cyclic universe, continually running down and continually rejuvenating itself, seems to me wholly retrograde" (Eddington 1935: 59; emphasis added). Eddington was not the only astronomer to feel in this way. In a lecture of 1940, the distinguished American astrophysicist Henry Norris Russell expressed his surprise of "the wide-spread desire to believe in some cyclical restoration of however great intervals" (Russell 1940: 27). With regard to this question, which he considered to be aesthetically rather than religiously based, he sided with Eddington.

　　Although short-lived and merely a more elaborate version of what Friedmann had shown earlier, Einstein's pulsating model attracted considerable attention, both scientifically and among a broader audience. In May 1931 Einstein went to Oxford to receive an honorary doctorate and give a series of three Rhodes Memorial Lectures. The second of the lectures, delivered on 16 May, dealt with the "cosmologic problem" and included a discussion of his as yet unpublished model of the cyclic universe without a cosmological constant.[10] In a book titled *God and the Astronomers* the theologian William Ralph Inge, Dean of St. Paul's, commented on Einstein's view "that the ponderable matter of the universe alternately expands and contracts" from a religious perspective: "This, if it is Einstein's settled view, is a revolutionary change, for it means a return to the old theory of cosmic cycles, which has long attracted me" (Inge 1934: 50). Contrary to most other theologians and Christian thinkers, Inge subscribed to the view that an eternally oscillating universe was in full agreement with Christian thought. Einstein's model was not actually perpetually cyclic, but Inge evidently thought it was.

---

[10]　The lecture was not published, but a brief summary of it appeared in *Nature* 127 (1931): 790. The blackboard Einstein used, filled with his calculations of the cyclic universe, was kept and can be seen at the Museum of the History of Science in Oxford. The formulae on the blackboard correspond closely to those in his published paper.

In an important paper of 1933, Georges Lemaître studied the "annihilation of space," that is, the problematic singularity where "the radius of space may pass through zero."[11] In this connection he used as an example what he called "Einstein's cycloidal universe." However, he concluded that the model could not be correct because it was unable to provide an age of the universe (the time since the initial singularity) longer than 1-2 billion years. Lemaître tended to regret the non-physical nature of oscillating solutions "from a pure aesthetic point of view," which he explained as follows: "Those solutions where the universe expands and contracts successively while periodically reducing itself to an atomic mass of the dimensions of the solar system, have an indisputable charm and make one think of the Phoenix of legend" (Lemaître 1997: 679; Lambert 1999: 152-155).

Many years later, in his presentation to the Solvay Congress of 1958, he returned to the Phoenix universe in which "any detail of the contraction period should have been destroyed." He argued that the new expansion would result in a mass of gas in a state of maximum entropy. Keeping to his old idea of an exploding primordial atom as the source of the expanding universe, he said (Lemaître 1958: 9):

> On the contrary, the distribution coming out from fresh matter would be a distribution of minimum entropy, i.e. a very unprobable distribution, very far from thermodynamic equilibrium. … The only feature it has in common with a gas is that it is formed of a great number of individual "molecules," but they have not the Maxwellian distribution which is the real characteristic of a gas. It is better to describe such a state of matter as being corspuscular radiation travelling along in every direction. … I do not see how a useful cosmology can be built by starting from the Phenix nucleon gas.

It is sometimes stated that Lemaître was in favor of the Phoenix universe model, but this is not the case. He discussed it briefly at a few occasions, without ever advocating it. Indeed, being a Catholic priest it would have been most surprising if he had adopted such a picture of the universe, traditionally being associated with materialism and atheism.

During the 1930s oscillating models of the type first described by Friedmann were well known and included in the reviews of Robertson, Otto Heckmann,

---

[11]  Lemaître 1933 was reprinted in 1972 in *Pontifical Academiae Scientiarum Scripta Varia* 36: 107-181. In 1997 an English translation by M. A. H. MacCallum appeared in *General Relativity and Gravitation* (Lemaître 1997).

Zaycoff, George McVittie and others. They entered the scientific literature alongside other relativistic models, but were rarely seen as particularly important or interesting. To my knowledge, no physicist or astronomer in the period expressed strong commitment to the idea, whereas a few, such as Eddington and de Sitter, reacted emotionally against it.

## 3. Richard Tolman and cosmic entropy

A thorough investigation of cyclic models was first undertaken by Tolman, the eminent and versatile American physicist who since 1922 worked as professor at the California Institute of Technology (Caltech). Tolman was a pioneer in applying general relativity and thermodynamics to the universe at large. For example, as early as 1922 he applied chemical equilibrium theory to the hypothetical formation of helium from hydrogen in a static Einstein universe, concluding that the relative abundances of the two elements could not be explained in this way (Tolman 1922).

In later works Tolman derived expressions for the total energy and entropy of the universe, first in the static case and next for models of the evolving universe. He was the first to express the second law of thermodynamics in a covariant formulation and to discuss the cosmological consequences of it (Tolman 1931a).[12] He originally examined an expanding radiation-filled universe, in which case he concluded that periodic solutions could not occur. However, he also emphasized that "we must not conclude therefrom that periodic solutions would be of no interest for the actual universe" (Tolman 1931a: 1660).

A main point of his investigations from the early 1930s was the demonstration that if the relativistic form of thermodynamics was applied to the universe it would lead to results very different from those based on the classical thermodynamic reasoning of the late nineteenth century. Significantly, there would be no justification for either entropic creation or the heat death, that is, the creation of the world in a low-entropy state or the end of it in a state of maximum entropy. In his important textbook of 1934, *Relativity, Thermodynamics and Cosmology*, Tolman formulated this general conclusion as follows: "It would seem wisest, if we no longer dogmatically assert that the principles of thermodynamics necessarily require a universe which was created at a finite time in the past and which is fated for stagnation and death in the future" (Tolman 1934: 444).

---

[12] Relativistic thermodynamics started much earlier. For a critical survey of the early development, see Liu 1992.

Einstein was well aware of Tolman's works, which he admired. He corresponded with him in the 1930s, and in early 1931, when he spent a couple of months in Pasadena, he had intensive discussions with the American theorist (Nussbaumer & Bieri 2009: 145-148). Einstein's paper on the pulsating model, submitted to the Prussian Academy of Sciences on 16 April after his return to Berlin, was probably indebted to his interaction with Tolman. Einstein showed his appreciation of Tolman's contributions in a laudatory review he wrote of *Relativity, Thermodynamics and Cosmology* in 1934. In the review, which was published in *Science* and appeared in the original German (!), he explained his reasons for having abandoned the cosmological constant.[13]

According to Tolman, whereas a cyclic universe contradicted the classical version of the second law of thermodynamics, in relativity theory processes could take place without any increase in entropy at all. A world model such as Einstein's periodic universe could expand and contract reversibly without increase in entropy, he concluded (Tolman 1931b). Moreover, extending his analysis to systems with irreversible processes he found that it was possible for a closed universe to undergo a continual series of cycles if only $\Lambda \leq 0$. In general, "relativistic thermodynamics could not impose restrictions which would prevent such a series of expansions and contractions" (Tolman 1932a: 331).

Tolman's analysis showed that the simple picture of identical cycles had to be replaced with one in which each new cycle became greater than the previous cycle, both with respect to the period and the maximum value of the curvature radius. As to the entropy, although it would increase from one cycle to the next it would never attain or approach a limit of maximum entropy. Alluding to the discussion in the nineteenth century – or perhaps to Eddington's advocacy of the heat death scenario – he concluded that a succession of expansions and contractions could occur "without ever coming to that dreadful final state of quiescence predicted by the classical thermodynamics" (Tolman 1932b: 372).

The models proposed by Tolman continued to inspire cosmologists many years after his death in 1948 (Heller & Szydlowski 1983). For example, Peter Landsberg and David Park (1975) examined the entropy in an oscillating universe model by means of computer experiments.[14] Their results were in broad agreement with those found by Tolman, including that successive cycles become larger and larger. They also found, again in agreement with Tolman, that the entropy increases continuously and can thus, even in an oscillating universe, be used as a measure for the direction of time. These results were confirmed by Landsberg and Reeves (1982), who further showed that the model collapses faster than it expands, that is,

$$\left|\frac{dR}{dt}\right|_{final} > \left|\frac{dR}{dt}\right|_{initial}$$

---

[13] *Science* 80 (1934): 358. A more visible result of Einstein's stay in Pasadena was a brief paper he wrote jointly with Tolman and Boris Podolsky on the philosophical problems of quantum mechanics (Einstein, Tolman & Podolsky 1931).

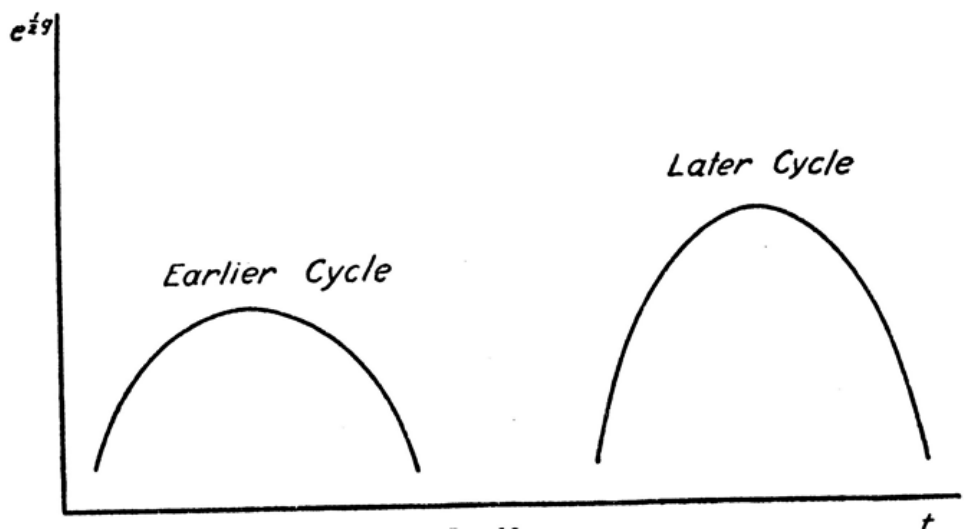

Fig.1. Tolman's illustration of two cycles of the oscillating universe, with the later cycle being greater than the earlier one. The quantity $e^{g/2}$ represents the radius of the curvature (Tolman 1934: 443).

Tolman's analysis of a pulsating universe filled with matter and radiation was valid for what he called quasi-periodic models, corresponding to an expansion of the universe from $R = 0$ to an upper limit and followed by a contraction to the same singular state. "Strictly periodic solutions" were outside the mathematical analysis, but Tolman thought that such a continual series of successive expansions and contractions might well be possible, indeed highly probable, from a physical point of view. He convinced himself that the conception of an ever-oscillating universe was not only "conceivable" but also "reasonable." According to him, in a physically realistic universe supposed to be not strictly isotropic and homogeneous, the singularity would not appear.

---

[14] All considerations of the entropy of the universe, whether supposed to be closed or not, rest on the assumption that the idea of entropy can be applied to the universe as a whole. It was and still is rarely realized that this is a questionable assumption. According to Robert Wald, there is "no reason to expect that there will be a meaningful notion of the 'total entropy of the universe'" (Wald 2006: 396).

Although Tolman was unable to provide a plausible physical mechanism for the bounce that supposedly led to a new cycle, he suggested that such passages through $R = 0$ (or $R = R_{min}$) were "physically inevitably necessary" (Tolman 1934: 439). In lack of a mechanism he suggested an analogy to the behavior of an elastic ball bouncing up and down from the floor: Newton's second law in the form $d^2x/d^2t = -g$ can describe how the ball rises to a maximum height and subsequently falls to the floor, but it cannot describe the mechanism of reversal when the ball hits the floor. Considerations concerning the elastic properties of the ball and the floor have to be taken into account, and these are not provided by the equation of motion.

Tolman was well aware of the danger of confusing a cosmological model with the real universe, something he often warned against. Yet, in spite of his cautious attitude he seems to have believed that his analysis justified an eternally oscillating universe rather than an explosive universe of Lemaître's type. He found it "difficult to escape the feeling that the time span for the phenomena of the universe might be most appropriately taken as extending from minus infinity in the past to plus infinity in the future" (Tolman 1934: 486).

In an important paper coauthored by his student Morgan Ward he derived from Einstein's field equations an early version of the singularity theorem, namely that a contracting closed universe with $\Lambda = 0$ will end in a zero volume.[15] He nonetheless believed that "from a physical point of view … we might expect contraction to the lower limit to be followed by a renewed expansion" and spoke in favor of the "possibility that the actual universe or parts thereof might also exhibit such a continued succession of expansions and contractions" (Tolman & Ward 1932: 837, 843). It is unclear if he realized that there can only be a limited number of preceding cycles, such as follows from the increase in entropy from cycle to cycle. If he did, he did not mention it.

Although Tolman certainly found oscillating models to be scientifically attractive, his interest in them did not extend to an emotional or philosophical commitment. In early 1932 he gave an address to the Philosophical Union, a society at the University of California at Los Angeles, in which he discussed the new cosmological models and his own work on the thermodynamics of the universe. "In studying the problem of cosmology," he said, "we are immediately aware that the future fate of man is involved in the issue, and we must hence be particularly careful to keep our judgments uninfected by the demands of religion, and unswerved by human hopes and fears." He continued (Tolman 1932b: 373):

> Thus, for example, what appears now to be the mathematical possibility for a highly idealized conceptual model, which would expand and contract without ever coming to a final state of rest, must not be mistaken for an assertion as to the properties of the actual universe, concerning which we still know all too little. … Although I believe it is appropriate to approach the problems of cosmology with feelings of awe for their vastness and of exultation for the temerity of the human spirit in attempting their solution, they must also be approached at the same time with the keen, balanced, critical and skeptical objectivity of the scientist.

---

[15] For an in-depth analysis of the history of singularity theorems, see Earman 1999.

A similar statement appeared in his book of 1934, where he warned against a realistic interpretation not only of Lemaître's new primeval-atom model but also of cyclic models (Tolman 1934: 488). Tolman's lack of strong commitment to an oscillating universe is confirmed by his later publications in cosmology, where this kind of model was in no way highlighted. For example, in a survey article of 1937 he merely mentioned the oscillating model as one possibility among others, and in his last paper on cosmology, published posthumously, he did not mention it at all (Tolman 1937: 37; Tolman 1949).

The oscillating model was not only problematic from a theoretical point of view but also because it was confronted by observational problems. For one thing, it shared with the Einstein-de Sitter open model the problem of an age of the universe that was shorter than the age of the stars and even of the Earth. In addition, it required a high average density of matter to reverse the motion of the universe. In a closed universe with $\Lambda = 0$ the density must exceed the critical value given by

$$\rho_{\mathrm{crit}} = \frac{3H_0^2}{8\pi G}$$

With the accepted value of the Hubble constant $H_0 \cong 500$ km/s/Mpc this meant a density greater than $10^{-28}$ g/cm$^3$. This was indeed a problem, but not a fatal one. In *The Realm of the Nebulae* of 1936, Hubble inferred from observations $10^{-30}$ g/cm$^3$ as a lower limit and $10^{-28}$ g/cm$^3$ as an upper limit. Other astronomers in the 1930s were willing to accept a mean density as high as $10^{-27}$ g/cm$^3$. Given the state of uncertainty in the observations, a cyclic universe remained a possibility, if not perhaps a very likely one.

It should be noted that not all cyclic conceptions of the universe in the interwar period were based on Einstein's general theory of relativity. The older conception of an eternally regenerating universe in the style of the nineteenth century continued to be discussed, quite independent of the oscillating models based on relativistic cosmology. Although these ideas were separate from mainstream cosmology and ignored by most physicists and astronomers, they enjoyed a considerable public support and were advocated by a few scientists of distinction (Jaki 1979: 342-345; Kragh 1995).

Hypotheses of a continually recycling universe with an equilibrium between organization and dissipation processes (or processes consuming and producing entropy) were suggested by, among others, Emil Wiechert and Walther Nernst in Germany, Robert Millikan and William MacMillan in the United States, and Oliver Lodge in England. These ideas had in common that they postulated an eternal universe that was perpetually creative without ever approaching a heat death. Accepting the traditional notion of space being flat and static, they were not cosmological models in the sense of relativity theory. They were recycling or regenerating world pictures, not cyclic in the sense of exhibiting a temporal periodicity of space over long spans of time.

## 4. The oscillating universe in the 1950s

Much of the development in cosmology during the first two decades following World War II was concerned with the controversy between relativistic evolutionary models and the rival steady-state theory introduced by Fred Hoyle, Thomas Gold and Hermann Bondi in 1948 (Kragh 1996). A cyclic universe obviously contradicted the basis of the steady-state theory, the perfect cosmological principle, whereas it was compatible with the idea of an exploding universe such as developed by George Gamow and his collaborators Ralph Alpher and Robert Herman in the late 1940s. On the other hand, Gamow's research program focused on the very early universe shortly after the big bang, whereas the geometry and long-time behavior of the universe was thought to be of less importance.

Gamow speculated that the expanding universe was presumably the result of an earlier contraction, a picture not unlike the one de Sitter had considered two decades earlier. However, he also pointed out that "there is no sense in speaking about that 'prehistoric state' of the universe, since indeed during the state of maximum compression … no information could have been left from the earlier time if there ever was one" (Gamow 1951: 406). Still, the bouncing model appealed to him, such as he made it clear in his best-selling popular book *The Creation of the Universe*. He coined the name "big squeeze" – today often known as the "big crunch" – for the hypothetical collapse of the universe that might have preceded the expanding one we live in (Gamow 1952: 36-37). In a paper of 1954 he concluded that such a model was "much more satisfactory" than the finite-age explosion model of Lemaître. His picture was this (Gamow 1954: 63):

> Thus we conclude that our Universe has existed for an eternity of time, that until about five billion years ago it was collapsing uniformly from a state of infinite rarefaction; that five billion years ago it arrived at a state of maximum compression in which the density may have been as great as that of the particles packed in the nucleus of an atom …, and that the Universe is now on the rebound, dispersing irreversibly toward a state of infinite rarefaction.

Realizing that this was a speculation, he cautiously added that "from the physical point of view we must forget entirely about the pre-collapse." Whereas the theist Lemaître had postulated a primeval atom of nuclear density without explaining its origin (which he thought was divinely caused), the atheist Gamow suggested a similar picture by assuming that the primeval atom was not truly primeval.

As to the possibility of a periodic universe with either a finite or an infinite number of cycles, Gamow dismissed it as incompatible with observational data. His universe was infinite in size and, he asserted, "the distances between the neighboring galaxies are bound to increase beyond any limit." Consequently, "there is no chance that the present expansion will ever stop or turn into a collapse" (Gamow 1952: 42). He repeated the verdict in a popular paper of 1956, but with the reservation that there might be large amounts of dark matter in intergalactic space, in which case the universe would be of the oscillating type (Gamow 1956).

Some of the scientists advocating oscillating models in the 1950s and later were motivated by a wish to avoid the perplexing problem of an origin of the universe. They found a primordial state, whether in the form of a space-time singularity or a primeval atom in the sense of Lemaître, to be unacceptable from both an observational and a philosophical point of view. For example, this was the opinion of the French astrophysicist Alexandre Dauvillier, professor at the Collège de France, who believed that any notion of a finite-age universe was metaphysical and anthropomorphic. Those who entertained such ideas entered "an unintelligible metaphysical terrain," he wrote. "Not only is the hypothesis [of Lemaître] not justified by observations, but it is a priori inadmissible because of its metaphysical character. It implies a supernatural creation *ex nihilo*, which remains outside scientific thought" (Dauvillier 1963a: 76, 95). The objection was identical to the one raised by Hoyle and a few other steady-state proponents. The French physicist further claimed that the popularity of Lemaître's hypothesis was to a large extent due to its exploitation by writers and scientists of a mystical or religious orientation.

As an alternative he advocated an infinity of cosmic cycles, which he thought could provide a framework for understanding the cosmic rays and the formation of chemical elements (Dauvillier 1955; Dauvillier 1963a; Dauvillier 1963b). However, although he was a staunch supporter of *la théorie des cycles cosmiques*, it was not in the cosmological sense of general relativity but in the older and more restricted sense of energetic cycles occurring endlessly in the universe. Strangely, he did not refer to cyclic models in the tradition of Friedmann, Einstein and Tolman. Dauvillier's ideas were closer in spirit to the earlier ones of Nernst, Millikan and MacMillan than to the oscillating models based on relativistic cosmology.

In the 1950s, relativistic models of the oscillating universe were considered in particular by William Bowen Bonnor in England and Herman Zanstra in the Netherlands. Bonnor, a specialist in general relativity originally trained in physical chemistry, investigated in a series of works the problem of galaxy formation and its connection to cosmological theories. In this context he argued for an oscillating universe in which the inhomogeneities of the early universe, out of which seeds galaxies were formed, were fossils from the preceding contraction (Bonnor 1954; Bonnor 1957). Admitting that there was no known physical mechanism that could reverse the contraction, he argued that "appropriate pressure changes will cause the model to change from contraction to expansion without passing through a singular state" (Bonnor 1954: 20). He realized of course the old objection against the eternally oscillating universe based on the second law of thermodynamics, but thought that it scarcely deserved to be taken seriously. As he wrote in a popular book of 1960, "it has never been properly shown how the Second Law of Thermodynamics affects the universe as a whole" (Bonnor 1960: 10). For some years he continued advocating an ever-oscillating model with no singular states. This model, he said, "has some of the advantages of the steady-state universe, without the very serious theoretical disadvantages" (Bonnor 1964: 204).

Herman Zanstra is well known as a distinguished astronomer and astrophysicist, but his role as a cosmologist has remained unacknowledged. A specialist in the physics of gaseous and planetary nebulae, he served as professor of astronomy at the University of Amsterdam, and in 1961 he received the Gold Medal of the Royal Astronomical Society, arguably the most prestigious prize in the astronomy community (Plaskett 1974; Osterbrock 2001). Four years earlier, in a little known paper in the proceedings of the Royal Dutch Academy of Science, he examined in great detail oscillating models of the universe (Zanstra 1957). Although the work attracted very little attention, it deserves a place in the history of cosmological thought.[16]

Building on Tolman's earlier works, Zanstra concluded that the oscillating universe was allowed observationally if only Hubble's old distance scale was increased by a factor of five or more. Since observers at the 200-inch Hale Telescope at the Palomar Observatory had recently determined the Hubble time to be at least 5.4 billion years (Humason, Mayall & Sandage 1956), as compared to the earlier estimated 1.8 billion years, he thought that an oscillating universe was a possibility. As to the question of whether the contraction of the universe could be reversed at a very small radius, or alternatively would end in a singularity, Zanstra (1957: 114) concluded: "To stop the compression of the universe so that it can be followed by expansion would require a high negative pressure, which seems to be physically excluded." Still, the conclusion assumed that the known laws of physics, general relativity included, were valid at extremely high densities, and Zanstra was willing to question the assumption. In this regard he was in good company, for authorities such as Einstein and Tolman had advocated somewhat similar lines of thought (Einstein 1931: 235; Tolman 1934: 438-439; Earman 1999: 240-242).

Following Tolman's analysis and assuming a repetition of cycles, Zanstra argued that with each new cycle the universe would grow bigger, with greater values of the period and the maximum curvature radius. During the final phase of a contraction, near the minimum value of $R$, he found that the temperature would increase drastically and at $R = R_{min}$ the matter of the compressed universe would consist of a hot gas of electrons and protons. From this state of maximum compression a new cycle would start afresh. However, based on thermodynamic arguments he concluded that the oscillating universe could not have existed prior to a certain time, that is, it could only have been preceded by a finite number of cycles: "Since at each reversal point a substantial more or less fixed amount of radiation is added, there cannot have been a whole pulsation prior to the reversal point where the radiation is less than this fixed amount" (Zanstra 1957: 119). A similar result, based on the finite value of the entropy per baryon, became important in cosmology about a decade later (Kragh 2009).

---

[16]  Zanstra 1957 appeared separately as circular no. 11 of the Astronomical Institute of the University of Amsterdam. It is not included in the Web of Science (Thomson-Reuters) and has not received attention by historians of science. The paper was abstracted in *Astronomischer Jahresbericht* 57 (1957): 124-125 but not in the *Physics Abstracts*.

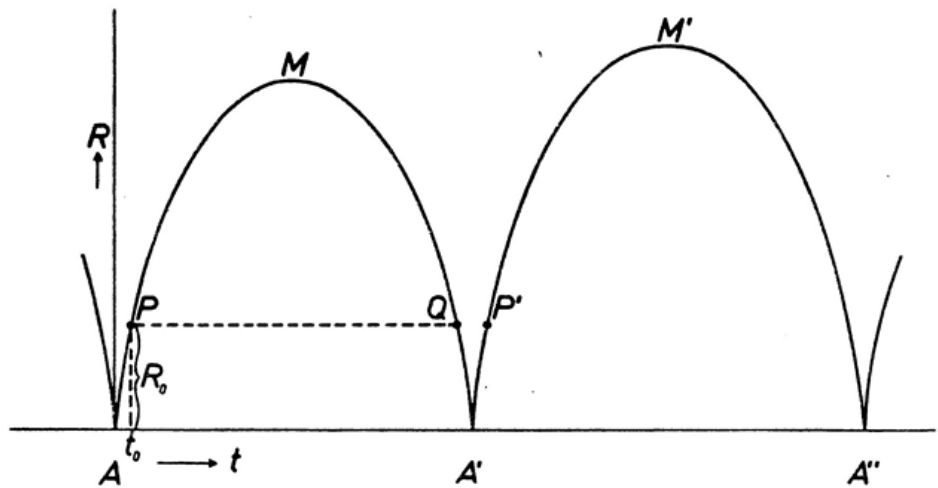

Fig. 2. The pulsating universe as depicted in Zanstra 1957: 116. The cycle starts in $A$ and bounces at maximum density ($A'$, $A''$) are assumed possible. The state of the present universe is represented by the point $P$, having radius $R_0$ and age $t_0$. $Q$ marks the later contracting state of the same radius. Notice that the second cycle $A'M'A''$ is greater than the first cycle $AMA'$.

In spite of his sympathy for the oscillating universe, Zanstra found that it did not quite live up to what he called the "philosophical desires" of how the universe should evolve. These desires he formulated in three principles:

(1) The universe must exist eternally, both in the past and in the future.
(2) The universe must be self-regenerating and never end up in a state of thermal equilibrium.
(3) Over long intervals of time, the universe must remain unchanged.

While ever-expanding models, such as of the Lemaître or Einstein-de Sitter type, violated all three principles, the oscillating model satisfied the second one: "Only one of the philosophical desires can be satisfied if reversal is assumed and none under ordinary laws" (Zanstra 1957: 121).[17] This made the cyclic universe appealing, but not appealing enough.

Zanstra returned to the oscillating universe in a paper of 1967 in which he reconsidered the problem of the finite number of past cycles that was responsible for the violation of the first of his philosophical desires. He now introduced the highly unorthodox idea of "a series of occult non-physical interventions at every compression" in order to maintain a cyclic universe with an eternal past existence (Zanstra 1967: 39). These occult forces he described as originating in a conscious spiritual reality, a divine being of some sort. His attempt to introduce spiritual philosophy in science was politely ignored by his fellow astronomers. Yet he meant it seriously and it was not the first time he suggested such ideas. In a series of lectures at the philosophy department of the University of Michigan 1959-1960 he expounded in some detail his metaphysical beliefs, which included telepathy and other parts of parapsychology (Zanstra 1962).

Interestingly, the revolutionary change in cosmology that occurred in 1965 with the recognition of the cosmic background radiation was in part motivated by the same idea that motivated Gamow's earlier work, namely, the oscillating or pulsating universe. Robert Dicke of Princeton University was in the early 1960s attracted to the idea, which led him to suggest the existence of a blackbody radiation left over from the last big crunch. According to his collaborator James Peebles, he was inspired by Tolman's demonstration that in a violent bounce entropy would be produced in the form of thermal electromagnetic radiation (Peebles, Page & Partridge 2009: 40-42, 286-188). It was Dicke's collaboration with Peebles, Peter Roll and David Wilkinson that resulted in the interpretation of the Penzias-Wilson microwave background at $\lambda = 7.35$ cm as a fossil from the big bang.

Although Dicke never published his speculations on the oscillating universe, they are visible in the seminal 1965 paper in *Astrophysical Journal* written by him and his three collaborators (Dicke, Peebles, Roll & Wilkinson 1965: 414):

---

[17]  It would seem that the steady state universe based on the perfect cosmological principle satisfied all three principles and thus was highly desirable, but Zanstra claimed that it failed on account of the third principle.

The matter we see about us now may represent the same baryon content of the previous expansion of a closed universe, oscillating for all time. This relieves us of the necessity of understanding the origin of matter at any finite time in the past. In this picture it is essential to suppose that at the time of maximum collapse the temperature of the universe would exceed $10^{10}$ °K, in order that the ashes of the previous cycle would have been reprocessed back to the hydrogen required for the stars in the next cycle.

A few lines later the authors admitted that "we need not limit the discussion to closed oscillating models." There is little doubt that the reference to the oscillating universe reflected Dicke's predilection for it.

## 5. Negative pressure as a saving device

While Zanstra dismissed a negative pressure as unphysical in his paper of 1957, in his later article in *Vistas in Astronomy* he referred to the hypothesis such as introduced by William McCrea (1951) and adapted to cyclical models by the Polish physicist Jaroslav Pachner (1965). As early as 1934 Lemaître had pointed out that according to general relativity the vacuum corresponds to an ideal fluid with pressure $p$ and energy density $\rho$ given by

$$p = -\rho c^2 \quad \text{and} \quad \rho = \Lambda c^2 / 8\pi G$$

In the language of later cosmology, the equation of state for such a fluid is given by $w = -1$.[18] In 1951 McCrea introduced the concept in a revised version of the steady state theory, but without assigning any direct physical effects to it (McCrea 1951; Kragh 1999). His conceptual innovation reappeared in a bouncing non-singular model that George McVittie proposed in 1952 and in which he universe contracted to a minimum value after which it would continually expand (McVittie 1951). However, it took several years until ideas of a negative pressure made an impact on cosmological thinking.

When negative pressure was incorporated into cosmology it resulted in a proliferation of models (Harrison 1967; Clifton & Barrow 2007), and with the emergence of inflation theory in the early 1980s, and later with the recognition of dark energy, the negative pressure associated with $\Lambda$ became almost fashionable. Among the models that made use of the notion in order to avoid a singular state there was a class of universes oscillating gently without bangs or crunches, of which Pachner's model of 1965 was an early example. I shall not deal with these post-1960s models (see Kragh 2009) except pointing out that the idea of a negative pressure helped oscillating models to survive at a time when few cosmologists considered them to be a viable alternative to the standard big-bang picture.

---

[18] Lemaître 1934. He actually stated the denominator as $4\pi G$ rather than $8\pi G$ and took $\Lambda > 0$ to correspond to a negative vacuum energy density (and a positive pressure density). In cosmology the equation of state is given by the dimensionless number $w = p/\rho$. For an ordinary gas $w = 0$, whereas for radiation and relativistic matter, as in the early universe, $w = 1/3$. It can be shown that for $w < -1/3$ the expansion of the universe is accelerating. One example is dark energy in the form of the cosmological constant, where $w = -1$.

The singularity theorems proved by Roger Penrose, Stephen Hawking and others in the mid-1960s demonstrated that cosmic singularities were nearly unavoidable and therefore raised serious doubts as to the possibility of a bounce from one cycle to the next (Earman 1999). On the other hand, the problem of cosmic singularities was not specifically related to oscillating models and none of the investigations that led to the singularity theorems mentioned these models as particularly problematic. Moreover, workers in the field realized that their arguments for singularities were not waterproof. Thus, in 1956 Arthur Komar showed that cosmological singularities are to be expected under very general assumptions, but he also noted that a negative pressure could prevent the occurrence of singular states (Komar 1956: 546). Hawking similarly referred to the negative-energy $C$ field in the steady-state theory of Hoyle and Jayant Narlikar as a possible way to avoid the cosmic singularity (Hawking 1966: 521).

With or without the hypothetical negative pressure, the oscillating universe faced serious problems of both an observational and theoretical nature. In spite of the problems, this class of models continued to attract attention and be investigated by a minority of cosmologists, eventually leading to the revival of interest in the twenty-first century referred to in the introduction.

*Acknowledgments.* I thank the Caltech Institute Archives for permission to quote from Tolman's unpublished correspondence. Tilman Sauer kindly provided me with copies of letters between Einstein and Tolman, for which I am grateful.